\documentclass[tradiabstract]{aa}

\usepackage{natbib}
\bibpunct{(}{)}{;}{a}{}{,}
\usepackage{graphicx}
\usepackage{txfonts}

\begin{document}

\title{A $Hubble\ Space\ Telescope$ transit light curve for GJ\,436b\thanks{Table~2 is only available in electronic form at the CDS via anonymous ftp to cdsarc.u-strasbg.fr (130.79.128.5) or via http://cdsweb.u-strasbg.fr/cgi-bin/qcat?J/A+A/}}

\author{J. L. Bean\inst{1} \and G. F. Benedict\inst{2} \and D. Charbonneau\inst{3} \and D. Homeier\inst{1} \and D. C. Taylor\inst{4} \and B. McArthur\inst{2} \and A. Seifahrt\inst{1} \and S. Dreizler\inst{1} \and A. Reiners\inst{1}}

\institute{Institut f\"{u}r Astrophysik, Georg-August-Universit\"{a}t G\"{o}ttingen, Friedrich-Hund-Platz 1, 37077 G\"{o}ttingen, Germany\\
\email{bean@astro.physik.uni-goettingen.de} 
\and McDonald Observatory, University of Texas, 1 University Station, C1402, Austin, TX 78712, USA
\and Harvard-Smithsonian Center for Astrophysics, 60 Garden Street, Cambridge, MA 02138, USA
\and Space Telescope Science Institute, 3700 San Martin Drive, Baltimore, MD 21218, USA}

\date{Received 21 April 2008; accepted 4 June 2008}

\abstract{We present time series photometry for six partial transits of GJ\,436b obtained with the Fine Guidance Sensor instrument on the \textit{Hubble Space Telescope} (\textit{HST}). Our analysis of these data yields independent estimates of the host star's radius $R_{\star} = 0.505^{+0.029}_{-0.020}\,R_{\sun}$, and the planet's orbital period $P = 2.643882^{+0.000060}_{-0.000058}$\,d, orbital inclination $i = 85.80\degr\ ^{+0.21\degr}_{-0.25\degr}$, mean central transit time $T_{c} = 2454455.279241^{+0.00026}_{-0.00025}$\,HJD, and radius $R_{p} = 4.90^{+0.45}_{-0.33}\,R_{\oplus}$. The radius we determine for the planet is larger than the previous findings from analyses of an infrared light curve obtained with the \textit{Spitzer Space Telescope}. Although this discrepancy has a 92\% formal significance (1.7$\sigma$), it might be indicative of systematic errors that still influence the analyses of even the highest-precision transit light curves. Comparisons of all the measured radii to theoretical models suggest that GJ\,436b has a H/He envelope of $\sim$ 10\% by mass. We point out the similarities in structure between this planet and Uranus and Neptune and discuss possible parallels between these planets' formation environments and dynamical evolution. We also find that the transit times for GJ\,436b are constant to within 10\,s over the 11 planetary orbits that the \textit{HST} data span. However, the ensemble of published values exhibits a long-term drift and our mean transit time is 128\,s later than that expected from the \textit{Spitzer} ephemeris. The sparseness of the currently available data hinders distinguishing between an error in the orbital period or perturbations arising from an additional object in the system as the cause of the apparent trend. Assuming the drift is due to an error in the orbital period we obtain an improved estimate for it of $P = 2.643904 \pm 0.000005$\,d. This value and our measured transit times will serve as important benchmarks in future studies of the GJ\,436 system.}

\keywords{techniques: photometric -- eclipses -- stars: individual: GJ\,436 -- planetary systems}

\maketitle

\section{Introduction}
Transiting exoplanets present a unique opportunity to evaluate planet formation and evolution models. The principle information obtainable from the study of such planets are their masses, radii, and orbits. This complete data set cannot be obtained for non-transiting planets using any other methods with current technology. The radius of a planet for a given mass and insolation is a strong constraint for theoretical models and yields information about the planet's structure. Additionally, carefully designed observations of transiting exoplanets can yield data that were, up until recently, widely thought to only be the purview of future direct imaging studies. Such data include, but are not limited to, exoplanets' chemical compositions, thermal emission, albedos, and energy redistributions \citep[for a recent review of the field see][]{charbonneau07}. 

The recent discovery by \citet{gillon07b} that the planet GJ\,436b, which was originally identified with Doppler spectroscopy by \citet{butler04}, transits its host star is an important advance in this rapidly burgeoning field. It is the first member of the emerging class of planets known as ``Hot Neptunes'' (i.e. planets having masses similar to that of Neptune and orbital semimajor axes $\sim$ 0.01 -- 0.10\,AU) that has been observed to transit. Therefore, detailed observational study of an ice giant outside our solar system can now be carried out.

Since the reported detection of transits for GJ\,436b there have been a flurry of follow-up observations and analyses of the system. So far, published results include complete analyses (i.e. data reduction and light curve modeling) by two separate groups of a single transit light curve and secondary eclipse light curve obtained at 8\,$\mu$m using the InfraRed Array Camera (IRAC) on the \textit{Spitzer Space Telescope} (hereafter \textit{Spitzer} for short). \citet[][hereafter G07]{gillon07a} and \citet[][hereafter D07]{deming07} analyzed the transit light curve and determined a precise mass, radius, and orbit for the planet. Both groups obtained consistent results and both concluded that the planet must have a gaseous H/He envelope similar to Neptune based on comparisons of its radius with theoretical models. Additionally, \citet[][hereafter S08]{southworth08} has recently analyzed the reduced \textit{Spitzer} light curve presented by G07 to give yet another estimate of the planet's radius, but did not use the result to make deductions about the planet's structure. The radius of GJ\,436b given by S08 is smaller than that ascertained by G07 and D07, although the results are consistent within the quoted uncertainties.

D07 and \citet{demory07} carried out complete analyses of the \textit{Spitzer} secondary eclipse light curve for GJ\,436b and were able to measure the planet-to-star flux ratio, and thus deduce the brightness temperature of the planet at the bandpass of the observations. The two groups found nearly identical results, but differed in their interpretation. D07 suggested that the measured temperature was higher than expected from a simple model of the planet and was indicative of tidal heating arising from the planet's non-circular orbit. Comparing the measured brightness temperature to more sophisticated models, \citet{demory07} found that a model with inefficient redistribution of energy from the day to night side could reproduce the observations. 

In this paper we present an analysis of a transit light curve for GJ\,436b obtained with the Fine Guidance Sensor (FGS) instrument on the \textit{Hubble Space Telescope} (\textit{HST}). Following the announcement of transits for GJ\,436b we were granted Director's Discretionary time to carry out these observations. The motivation was to use the exceptional photometric capabilities of the \textit{HST} \citep[e.g.][]{benedict98} to make a precise, independent determination of the radius and orbital ephemeris for the planet. Such measurements are of critical importance because they are the foundation on which the illuminating follow-up studies that are only possible for transiting planets are based. In the case of GJ\,436b specifically, an independent measurement of the radius is a test of previous assertions that the planet has a gaseous H/He envelope like Neptune and allows robust interpretation of secondary eclipse measurements. Additionally, the measured ephemeris will be a benchmark for future transit timing studies aimed at detecting additional planets in the system.

\section{Observations}
Six observations of partial transits for GJ\,436b were carried out between 4 December 2007 and 5 January 2008 using the FGS1r on \textit{HST}. This period spanned 11 orbits of GJ\,436b. A log of the observations is given in Table~\ref{t1}. The FGSs are white light shearing interferometers that use four photomultiplier tubes (PMTs), two in each orthogonal direction on the sky, to precisely measure tilts in the incoming wavefront. The measured counts from the PMTs are continuously recorded at a frequency of 40\,Hz (25\,msec). The instrument can be used as a high-speed photometer because of this, in addition to its more recognized uses for high precision relative astrometry and high angular resolution sensing. 

The FGS1r was used in ``Position'' mode for our observations. In this mode the instrument uses the measurements of the wavefront tilt to lock on to the observed photocenter and hold fixed pointing. At a given time the FGS measures the photometric signal in a selectable 5\arcsec x 5\arcsec\ sub-field out of the greater instrument field of view (FOV). Position mode observations do not allow simultaneous measurements of multiple targets. Therefore, observations of multiple targets must instead be carried out for each one separately.

We used the F583W filter for all observations. This filter has a nearly constant 95\% transmission function over the range 450 -- 700\,nm, and drops sharply to 0\% outside that range. The sensitivity of the PMTs decreases linearly from a maximum of 15\% to 4\% towards redder wavelengths over the non-zero transmission range of the filter. 

The observations consisted of six visits with a single orbit each. During each orbit we observed two reference stars with similar brightnesses (GSC 01984-02050 and GSC 01984-02056) as GJ\,436 before and after the main observation of it to facilitate correcting for inter- and intra-orbit variances in the photometry. The length of each reference star observation was 100\,s. The main observation of GJ\,436 in each orbit was 1800\,s. The observing sequence for all visits was: reference star \#1, reference star \#2, GJ\,436, reference star \#2, and then reference star \#1. 

\begin{table}
\caption{Log of GJ\,436 observations.}
\label{t1}
\centering
\begin{tabular}{rcc}
\hline\hline
UT date & Orbit phase ($T_{c}$) & Relative flux level \\
\hline
 4 Dec 2007  & 0.9942 -- 0.0021 & 0.99962 $\pm$ 0.00023  \\
 7 Dec 2007  & 0.0016 -- 0.0095 & 0.99967 $\pm$ 0.00017  \\
12 Dec 2007  & 0.9917 -- 0.9996 & 0.99906 $\pm$ 0.00021  \\
28 Dec 2007  & 0.9866 -- 0.9945 & 1.00358 $\pm$ 0.00011  \\
 3 Jan 2008  & 0.0015 -- 0.0094 & 0.99782 $\pm$ 0.00021  \\
 5 Jan 2008  & 0.9838 -- 0.9918 & 1.00024 $\pm$ 0.00010  \\
\hline
\hline
\end{tabular}
\end{table}

Initial processing of the raw data files was done with the FGS data reduction pipeline. This yielded the measured counts from each of the four PMTs, the calculated position of the target in the FOV, and data quality flags for each individual 40\,Hz sample. We retained readings obtained only when the instrument was in fine lock on one of the targets and no errors were indicated.

The pipeline also output the average counts per 40\,Hz sample for each of the PMTs when moving the instantaneous FOV from one target to the next during an orbit. This is a good estimate of the total dark and background counts for the PMTs. The estimated values for all four PMTs during all six visits agreed with the results from the nominal values given by \citet{fgsdata}. The values were on average $\sim$ 0.2\% of the measured counts for GJ\,436. The first step of post-pipeline processing was then to subtract the corresponding background from each 40\,Hz sample measured by the four PMTs.

After the background subtraction, we corrected each sample for the dead time in the PMTs according to the prescription of \citet{fgsdata}. We converted the geocentric observation times to the heliocentric frame using the recorded position of the spacecraft relative to the Sun and the observation vector to GJ\,436. The correction from the spacecraft frame to the geocentric frame was negligible. 

During observations in fine lock on a target the FGS continuously corrects the fine pointing using the measured count differences in the two orthogonally oriented PMT pairs. In an ideal instrument, this difference would be zero when the instrument was pointed directly at a target. However, the PMTs have different sensitivities and a parallel wavefront will result in small, but important, differences in PMT pair counts. This wouldn't be a problem for photometry in and of itself, except that the fine pointing of the instrument is not perfect and the apparent position of the star varies randomly. The impact of these two issues is that neither the average nor the total of the PMT counts in each pair is a conserved quantity. To correct for this we estimated the relative sensitivities of the PMTs in each pair using the average ratio of the counts recorded during the main observation of GJ\,436. This was done separately for each visit. We corrected the counts recorded by each of the low sensitivity PMTs in the two pairs for all observations in a single visit using this determined ratio. We then took the mean of the 40\,Hz samples from the two PMTs in each pair. This procedure gave two high frequency time series, one for the ``X'' axis and one for the ``Y'' axis, for each visit. The following steps were done for each of these two sets independently.

We attempted to correct for the inter- and intra-orbit variances in the photometry using the measurements for the reference stars. We first fitted a time dependent linear change in instrument response to the two observations of each reference star. The direction of change was always the same for each reference star within a single visit, but not the same for all the visits. The slopes for the two reference stars were averaged and the resulting linear trend divided from the GJ\,436 observations. The magnitude of this correction was less than 0.05\% for 90\% of the data points and always less than 0.1\%. We repeated the analysis described in \S 4 on data reduced using the correction from only on one or the other reference star and with no correction. In all the cases the determined parameters were indistinguishable from those we present below. Therefore, our results are robust againsit uncertainties in this step. The correction to the GJ\,436 photometry with the linear trend determined from the average of both reference stars yielded marginally smaller residuals from the light curve fit so it was utilized in the primary reduction.

The next step was to normalize the GJ\,436 data to the same relative flux scale. We were not successful at this no matter which combination of reference star data we used and the differences in the relative flux levels between visits were obvious by eye. We believe that the primary reason for this is due to the varying position of GJ\,436 in the instrument's FOV. This happened because the two gyro guiding mode for the \textit{HST} prevented the same spacecraft roll angle to be used for all the visits. The visit groups 1, 2, and 3 -- 6 were each obtained at different rolls. The FGS FOV must not have a flat photometric response at the level our data are sensitive to in addition to the well studied variance in position aberration \citep[i.e. the optical field angle distortion,][]{mcarthur02}. 

Examination of the un-normalized data for the multi-visit set obtained at the same roll show consistent, but still not perfect, relative flux levels. This supports our hypothesis about the FOV variance. The lower level disagreement for this group could be a result of still small variances in FOV position even among visits carried out with the same roll or the effects of stellar activity \citep{demory07}. We see no way to distinguish between the these two effects with the current data.

The reference stars were never placed in the same position in the FOV as GJ\,436, and indeed this would not have been possible due to \textit{HST} guiding and FGS pointing restrictions. Therefore, the reference star data cannot be used to correct the relative flux levels of GJ\,436. The time dependent response correction described above is still valid because that variance is most likely due to the thermal settling of the telescope itself and should be similar for all the targets. Our solution to the relative flux correction problem was to introduce normalization parameters for the data obtained in each of the visits that were solved for during the light curve analysis described in \S 3. The relative flux levels for the visits determined from this analysis are given in Table~\ref{t1}.

At this point in the reduction we had two time series (X and Y axes) for each of the six visits. We summed the two sets to make a single time series and analyzed the data as described in \S 3. This analysis yielded very poor results. The residuals were much larger than expected from counting statistics and clearly correlated (trends and jumps) on $\sim$ 10\,minute timescales. Inspection of the data revealed that the source for most of the unusual noise was the X axis data. When the data from the two axes were analyzed separately, we found that X axis data had residuals twice as large as those from the Y axis despite nearly identical count rates. Furthermore, the Y axis transit model residuals do not exhibit obvious correlations like the X axis data. We don't have a definitive explanation for the lower quality of the X axis data. We note that since beginning science observations with the FGS1r in 2000 we have consistently (thousands of independent observations) obtained position residuals $\sim$35\% higher in X axis data compared to Y axis data when using the instrument for high-precision relative astrometry even though corrections determined from extensive calibration efforts are applied for this work \citep[e.g.][]{benedict07,bean07}. This discrepancy is similar in magnitude but opposite what was seen in data from FGS3 when it was used for science observations. The effect we see in the GJ\,436 photometry is likely related to this issue, but relatively larger possibly due to a lack of any sort of known applicable correction. We ultimately decided to set aside the X axis data because the expected $\sqrt{2}$ reduction in counting noise from including these data is more than negatively compensated by the larger errors introduced by using it.

\begin{figure}
\resizebox{\hsize}{!}{\includegraphics{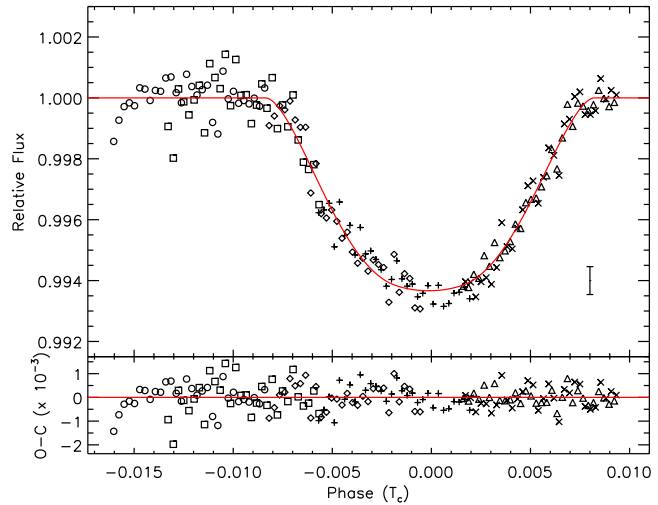}}
\caption{\textit{Top} Observed transit light curve for GJ\,436b (points) with best fit from our analysis (line) phased to the central transit time. The bar on the right indicates the average error size. \textit{Bottom} Residuals from the fit (points), which have a rms of 5.2 x 10$^{-4}$. The different point styles in both panels indicate data from the different visits.}
\label{f1}
\end{figure}

The final time series that we analyzed as described in \S 3 was created by binning only the Y axis 40\,Hz measurements to 60\,s samples, which yielded 180 data points. The adopted values for each bin were the average of the counts and the initial error estimate was the error in the mean. These data are given in Table~2, which is only available electronically from the CDS.
\setcounter{table}{2}

\section{Analysis}
\subsection{Parameter determination}
We modeled the obtained photometric time series of GJ\,436 using the exact analytic formulae given by \citet{mandel02} for a planetary transit. To account for the stellar limb darkening we calculated flux-weighted theoretical spectra for 18 different angles from the central line of sight and integrated them over the unique bandpass of the FGS with the F583W filter. We used the latest version of the PHOENIX model atmosphere code \citep{hauschildt99} for these calculations with the stellar parameters given by \citet{bean06} as determined from a spectral synthesis analysis ($T_{eff}$ = 3480\,K, log g = 4.92, and [M/H] = -0.33). We fitted the calculated specific intensities with the non-linear equation for stellar limb darkening proposed by \citet[][Eq. 6]{claret00} to obtain the necessary limb darkening coefficients for use with the \citet{mandel02} formulae ($c_1$ = 1.47, $c_2$ = -1.10, $c_3$ = 1.09, $c_4$ = -0.42).

We used a Markov Chain Monte Carlo (MCMC) method similar to that of \citet{holman06} to identify the best fit model to the data. In the analysis we fixed the values of the star's mass, the planet's (significantly non-zero) orbital eccentricity and longitude of periastron, and the radial velocity semiamplitude of the star due to the planet to those determined by \citet{maness07}. The parameters that we allowed to vary were the star's radius $R_{\star}$, the transiting planet's radius $R_{p}$, orbital inclination $i$, orbital period $P$, central transit time at a mean epoch of the observations $T_c$, and relative flux normalizations for each of the six visits. There were 11 free parameters in total. 

\begin{table}
\caption{Parameters for GJ\,436 and transiting planet.}
\label{t3}
\centering
\begin{tabular}{ll}
\hline\hline
\\[-5pt]
Parameter & Value \\[5pt]
\hline
\\[-5pt]
\textit{Determined directly in this paper} \\[5pt]
$R_{\star}$ ($R_{\sun}$) & $0.505^{+0.029}_{-0.020}$ \\[5pt]
$R_{p}$ ($R_{\oplus}$)   & $4.90^{+0.45}_{-0.33}$ \\[5pt]
$P$ (d)                  & $2.643882^{+0.000060}_{-0.000058}$ \\[5pt]
$i$ ($\degr$)            & $85.80^{+0.21}_{-0.25}$ \\[5pt]
$T_{c}$ (HJD)            & $2454455.279241^{+0.00026}_{-0.00025}$ \\[5pt]
\hline
\\[-5pt]
\textit{Inferred} \\[5pt]
$M_{p}$ ($M_{\oplus}$)   & $22.8^{+1.5}_{-1.5}$ \\[5pt]
$\rho$ (g\,cm$^{-3}$)     & $1.1^{+0.2}_{-0.3}$ \\[5pt]
\hline
\end{tabular}
\end{table}

We adopted the standard $\chi^{2}$ parameter as the fit quality metric. We initially used the uncertainties in the photometry as described in \S 2, but increased them by 18\% for the final analysis so that the best fit model had $\chi^{2}_{\nu}$ = 1.0. We combined the results from five Markov chains that were initialized with different randomly selected parameter values. The chains each contained 500\,000 points and the first 20\% were discarded to reduce the impact of the initial parameters. The perturbation sizes were adjusted to give a 25\% jump acceptance rate. The \citet{gelman92} $\sqrt{\hat{R}}$ statistic for the chains was within 1\% of unity for all the parameters, which indicates that the chain lengths were likely sufficient for convergence. We adopt the medians of the MCMC distributions as the determined parameter values. The uncertainties were estimated by the range of values that encompassed 68\% of the parameter distributions on each side of the corresponding median. The model for the adopted system parameters is shown with the observations phased to the central transit time in Fig.~\ref{f1}. The time series has a rms of 5.2 x 10$^{-4}$ in relative flux units, which corresponds to 0.6\,mmag.

To account for possible systematic errors in the determined parameters due to the uncertainties in the fixed parameters we repeated the MCMC analysis with each of these parameters adjusted to 1$\sigma$ higher and lower than their nominal values. We adopt the uncertainties in the fixed parameters given by \citet{maness07}. We found that only the uncertainty in the adopted stellar mass (0.04\,$M_{\sun}$) was a significant source of error in any of the parameters we determined. We added the parameter deviations from fixing the stellar mass to high and low values in quadrature with the uncertainties found from the MCMC analysis to give the total uncertainties. The determined parameter values and the inferred planet mass ($M_{p}$) and density ($\rho$) along with their corresponding uncertainties are given in Table~\ref{t3}.  

The central transit time determined from the above analysis and given in Table~\ref{t3} should be thought of as an epoch mean value because the observations to obtain the full light curve spanned 11 of GJ\,436b's orbits. We carried out an analysis of the visit data sets separately to determine the corresponding individual transit times. One of the visits occurred completely out of transit and was not included in this step. We fit a transit light curve model to the five visit data sets that are sensitive to the transit time using a Levenberg-Marquardt algorithm. We allowed only the time for each observed transit to vary and held the other parameters fixed to either their previously determined or adopted values. To estimate the uncertainties in the transit times we used a bootstrap Monte Carlo method. We generated 10\,000 realizations of each visit set by randomly drawing 30 data points with replacement. We also randomly modified one of the fixed parameters during each iteration by drawing from a normal distribution of values with a width set to the parameter's uncertainty. We modified only one of the fixed parameters for each iteration because the estimated errors in the parameters are correlated. We adopted the standard deviation in the distribution of times found from fitting the bootstrap trials as the uncertainty in the transit times. The determined times for each observed transit along with their uncertainties and deviation from the mean ephemeris are given in Table~\ref{t4}.

\begin{table}
\caption{Transit times and residuals from the mean ephemeris for GJ\,436b.}
\label{t4}
\centering
\begin{tabular}{cc}
\hline\hline
$T_{c}$ & O - C \\
(HJD)   & (d) \\
\hline
2454439.4160 $\pm$  0.0016 &  0.0001 \\
2454444.7036 $\pm$  0.0018 & -0.0001 \\
2454447.3477 $\pm$  0.0017 &  0.0001 \\
2454463.2109 $\pm$  0.0015 & -0.0000 \\
2454468.4986 $\pm$  0.0018 & -0.0001 \\
\hline
\hline
\end{tabular}
\end{table}

\subsection{Assessment of possible systematic error}
The necessity of fitting for the relative flux normalizations and phase folding partial transit observations in our analysis is unusual among the study of high-precision transit light curves. This raises the question of whether our results could be systematically wrong due to the inappropriateness of our approach. We investigated this possibility using a combination of re-analyses of the data with different assumptions and simulations.

To asses the validity of determining the physical parameters of GJ\,436 and its planet along with the flux normalizations we used a simulation technique. We generated ten simulated data sets with model parameters from the averaged results of G07, D07, and S08, random noise properties similar to our observed data, and random offsets in the visit subsets. We fitted these data using the MCMC procedure described above and examined whether there was a systematic deviation in the determined parameters compared to the known underlying parameters. We found no systematic trend in the deviations, and their distribution followed that expected from the adopted parameter uncertainties.

We also carried out a number of analyses on different realizations of the primary data set to ascertain whether some particular treatment of the data influenced our results. We repeated our analysis on subsets created by removing one of the visit sets to test the robustness of the results. We repeated this analysis six times, each time removing a different visit set. To ascertain whether the reduced time resolution from the 60\,s binning had an affect, we analyzed data sets where the 40\,Hz samples were binned to smaller ranges (10, 15, and 30\,s). We also repeated our analysis assuming a quadratic form of limb darkening (equivalent constants to the values for the non-linear model given above are $\gamma_1$ = 0.30, $\gamma_2$ = 0.49). Additionally, analyses were carried out by two of us separately and with slightly different methods. In all the cases the resulting parameters did not deviate from their nominal value given in Table~\ref{t3} by more than 1.5$\sigma$. 

Assuming a constant orbital period, transit timing deviations over the course of the observations, which spanned 11 orbits of the planet, could possibly adversely affect our analysis by distorting the morphology of the folded light curve. The transit timings we determined from our principle analysis are remarkably regular, but might not represent the true values as they depend on the physical parameters (i.e. stellar radius, and planet radius and inclination) determined from analyzing the phase folded light curve. To study this issue we repeated the transit timing analysis with the system parameters fixed to a representative of the G07, D07, and S08 determined values and allowing the relative flux normalizations to vary. From this we found possible timing variations of up to 79\,s and the $\chi^2$ for the fit was 170.9. For comparison the $\chi^2$ for our principle analysis was 168.6. for 169 degrees of freedom.

The low value of the fit quality metric from this alternative analysis suggests that the results from our primary analysis might not represent a unique model of the data. However, could true transit timings of this magnitude bias the system parameters determined by analyzing a phase folded light curve to the level our results are deviant from the \textit{Spitzer} results? To answer this question we generated simulated data similar to that described above for investigating the affect of variable flux normalizations. In this case we used transit times for the individual samples equal to those found from holding the system parameters fixed to those determined from the \textit{Spitzer} data. We then analyzed the simulated data in the same manner as we analyzed the real \textit{HST} data. From the analysis of 10 simulated data sets we found that transit timing offsets up to at least 80\,s did not significantly bias the results. For all the simulations the analysis algorithm returned the parameters that were used to create the simulated data within the one sigma uncertainties and without a systematic offset.

The above analysis suggests that transit timings have not caused our results to be biased. However, this analysis still poses the question of whether the inverse of the previous question is true. That is, can adopting systematically different physical parameters lead to spurious transit timing offsets? To study this we again turned to a simulation analysis. We generated simulated data using our determined system parameters and no transit timing offsets. We then repeated the analysis of fixing the parameters to those determined from the \textit{Spitzer} data while allowing the individual transit times and flux normalizations to vary. In this analysis we consistently found transit timing variations with the same magnitude ($\sim$\,80\,s) as those from above, and a fit $\chi^2$ close to, but still higher, than that obtained from using the correct (input) model parameters. As we know the transit timings are not correct because the simulated data were created assuming a constant period, we can be sure that transit timings determined from an analysis with assumed improper physical parameters would give spurious results.

Another possible source of uncertainty in our analysis is that arising from corellated, or ``red,'' noise \citep{pont06}. We assessed the influence of this type of noise on our results by re-fitting different realizations of the original data adjusted with the ``prayer bead'' method \citep{moutou04}. We used a Levenberg-Marquardt algorithm to perform the parameter optimization. The individual data sets were modified by shifting the residuals from the principle analysis described in \S 3.1 by a random number and adding them back to the data. This was repeated 10\,000 times. The standard deviation of the resulting parameter distributions gives the uncertainties. In this investigation we found uncertainties about half or less than those determined from the MCMC analysis. Thus, we conclude that corellated noise is not a significant source of uncertainty in the data. This is likely because the data set includes measurements from two different orbits (i.e. the two measurements are independent and uncorrelated) for almost every transit phase.

From all of the above investigations we conclude that our analysis is relatively unbiased, the determined parameters are a fair representation of the data, and the assigned parameter errors are realistic.

\section{Discussion}
\subsection{The planet's characteristics}
\begin{figure}
\resizebox{\hsize}{!}{\includegraphics{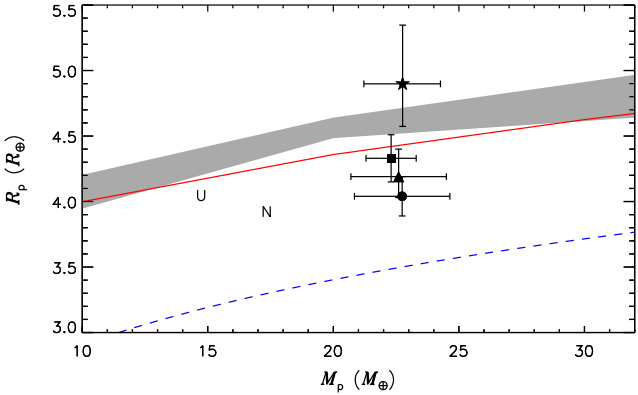}}
\caption{Determined mass and radius for GJ\,436b from G07 (triangle), D07 (square), S08 (circle), and this paper (star). The values for Uranus and Neptune are indicated by the ``U'' and ``N'' characters respectively. The lines are theoretical relationships for ice giants from \citet{fortney07}, where the solid line is for a planet with a 10\% by mass H/He envelope and the dotted line for a planet without an envelope. The shaded region delineates the possible range of values from \citet{baraffe08} models for an ice giant with a 10\% H/He envelope and age ranging from 1 -- 7\,Gyr. }
\label{f2}
\end{figure}

Our determined mass and radius for GJ\,436b is placed in context with the previous results of G07, D07, and S08, the solar system ice giants, and the theoretical mass-radius relationships of \citet{fortney07} and \citet{baraffe08} in Fig.~\ref{f2}. The radius we find differs by $1.82\sigma$ (93\% significance), $1.54\sigma$ (87\% significance), and $2.39\sigma$ (98\% significance) from the values of G07, D07, and S08 respectively. To collapse the previous results into a single value we first average the G07 and S08 results because they are from analyses of the same reduced data. Then we average this value with the D07 value to give $R_{p}$ = 4.22\,$R_{\oplus}$. We conservatively adopt a symmetric error in this value of 0.21\, $R_{\oplus}$, which is the largest error bar among the three results. The standard deviation of the three values is 0.15\,$R_{\oplus}$. The radius we find for GJ\,436b is larger by $1.74\sigma$ (92\% significance) than this representative of the previously determined values.

The \textit{Spitzer} data have an advantage in that the effect of limb darkening is almost non-existent due to the much longer wavelength of the bandpass used. However, we have carefully considered the effect of limb darkening in our analysis through specific calculation of it in the FGS bandpass using a realistic model atmosphere. Changing the parameters of the model atmosphere we used within reasonable limits or using a different functional form to fit the specific intensities (i.e. quadratic) does not significantly affect our results. Therefore, our determined parameters are relatively robust against uncertainties in the limb darkening. 

Our data have a similar level of precision and time sampling as the \textit{Spitzer} data. The larger uncertainties in the parameters from our analysis stems from the additional degrees of freedom introduced by fitting for the relative flux normalizations. However, as we have shown above, this does not mean the results are biased and is unlikely to be the reason for the planet radius discrepancy.

The larger radius we determine for the planet cannot construed as a detection of the planet's effective radius dependency on wavelength arising from spectral features \citep[e.g.][]{charbonneau02, barman07, pont08}, although our bandpass does cover the Na doublet near 600\,nm \citep{fortney03, barman07}. One reason for this is that we find a planet to star radii ratio consistent with G07, D07, and S08. If we fix the star's radius to the average of the values given by those authors we obtain a planet radius nearly exactly the same. Furthermore, our data are not sensitive to the likely size of the effect, which is probably at most a few percent in our wide bandpass.  

A plausible explanation for the differences between our results and the previous ones is stellar activity. As mentioned above, \citet{demory07} report that GJ\,436 exhibits visual flux variations with semiamplitude $\sim$ 1\%, and period $\sim$ 50\,d. These flux variations are consistent with periodic variations of emission in the Ca II H \& K lines, which presumably are modulated by the star's rotation.

Breaking the degeneracy between the planet and host star radii, as we have attempted to do, requires high-precision coverage of the transit ingress and egress \citep{charbonneau07}. It is possible that during the different visits the planet obscured regions with different brightnesses arising from the spots on the star's surface. If this happened during the ingress and/or egress then the morphology of the light curve was altered and our model was inadequate, which could result in incorrect determined parameters. It is also possible that if the planet obscured an unusually brighter or dimmer region of the star during the other parts of the transit then the fitting would be negatively affected due to the need to phase fold the partial transit observations and determine the relative flux normalizations. 

Another possible explanation for the discrepancy is that we, G07, D07, and S08 have simply underestimated our uncertainties. In addition to analyzing the \textit{Spitzer} light curve for GJ\,436b, S08 presented an interesting study of light curves for other transiting planet systems. One conclusion he reached was that analyses of even the highest-precision data sets for a given system can yield significantly different parameters owing to unidentified systematic errors in the data. In this context it is important to keep in mind that the G07, D07, and S08 results are not independent as they utilize the same data, albeit reduced independently by two of the groups and analyzed separately by all three. If there is a systematic error inherent to the \textit{Spitzer} data then the three groups would have obtained still similar, but systematically incorrect parameters. Formally, the difference between our planet radius and the representative \textit{Spitzer} value is significant. However, it could be that this difference is only representative of the true uncertainty in the analyses.

In the absence of any hard evidence that our result is spurious, we suggest that the radius of GJ\,436b is larger than previously thought. If we assume that a planet's radius dependence on the mass fraction of its envelope is linear in this regime, then extrapolation of the \citet{fortney07} models to our measured radius gives a H/He envelope mass fraction of 15\%. The weighted average of our radius value and the representative value of the \textit{Spitzer} results is 4.48 $\pm$ 0.17\,$R_{\oplus}$, for which the \citet{fortney07} models suggests an envelope mass fraction of 10\% $\pm$ 3\%. The \citet{baraffe08} models are fairly consistent with those of \citet{fortney07}, although they imply a slightly lower envelope mass fraction. \citet{baraffe08} also emphasize the impact of the evolutionary cooling on the radius of a planet like GJ\,436b and this adds to the ambiguity of specific interpretations because the age of GJ\,436 is quite uncertain. More theoretical and observational work is obviously needed to improve our understanding of GJ\,436b and, by extension, general models of planet formation and evolution.

From an observational standpoint, obtaining and analyzing an additional high-precision transit light curve of GJ\,436b with a different instrument than already used would still be valuable. Such data would preferably be obtained in the near-infrared to reduce the impact of stellar variability and uncertainties in the limb darkening. Another interesting observational study would be to make a direct radius measurement of the host star with interferometry, similar to what has been done for the transiting planet host star HD 189733 \citep{baines07}. We have found a similar ratio of the planet and star radii as the previous studies, and so an additional constraint on the stellar radius would give a tighter constraint on the planet radius. 

What can be definitively ascertained from comparison of the observational results and theoretical models is that GJ\,436b has a significant H/He envelope similar to Uranus and Neptune. Such similarities in structure suggest that they formed in a similar environment. Additionally, GJ\,436b and the solar system ice giants are similar in that they likely could not have formed in their current locations. Uranus and Neptune are thought to be too far away from the Sun \citep{levison01}, while GJ\,436b is likely too close to its host star. One hypothesis for the formation and evolution of Uranus and Neptune is that they were originally in the same region of the protoplanetary disk as Jupiter and Saturn \citep{thommes99}. The later two planets accreted gas much more quickly and scattered the ice giants to-be outward towards their current locations, which limited their growth. Uranus and Neptune are therefore really gas giant planet cores that did not accrete gas quickly enough to complete formation. In contrast, the structure of GJ\,436b likely does not require a comparably violent past because it orbits a M dwarf. \citet{laughlin04} have shown that formation of gas giants around low mass stars is severely hindered because these stars are expected to have correspondingly low surface density protoplanetary disks and the dynamical timescale of orbiting bodies is longer. \citet{laughlin04} predict a dearth of Jovian mass planets due to the resulting slow accretion of gas, but plenty of Neptune and lower mass planets around M dwarfs. In this context GJ\,436b could be considered a failed gas giant akin to Uranus and Neptune.

Furthermore, GJ\,436b could have experienced scattering due to another body like Uranus and Neptune even though this probably wasn't required to limit its growth into a larger body. The evidence for this comes from its current location, which is far away from its likely formation site; and its observed orbital eccentricity ($e$ = 0.16), which is in direct contradiction to predictions of tidal circularization theories \citep{maness07}. A major dynamical interaction event after formation leading to inward scattering is an explanation that unifies both of these properties into a single evolutionary picture. This interaction could have occurred with the outer object in the system that is indicated by the long term trend in GJ\,436's radial velocities. An alternative explanation for GJ\,436b's orbital properties is disk interaction leading to migration and eccentricity excitation \citep[e.g.][]{goldreich80, goldreich03}. Continued observations of the GJ\,436 system are needed to detect and characterize all of the objects it contains. Complementary theoretical studies based on the system census and the likelihood that GJ\,436 had a low-mass disk are also needed to specifically assess the plausibility of the conceivable evolutionary scenarios.

\subsection{Transit timings}
The results of our analysis also allow us to search for deviations in transit times that might arise from the gravitational perturbations of another planet in the system. The times determined for each of the five \textit{HST} partial transit observations, which are given in Table~\ref{t4}, exhibit a maximum deviation from the mean ephemeris of 10\,s. The average uncertainty on these times is 145\,s so the deviations are fully consistent with regular transit times over this observational period. 

To search for possible long-term variations we compare our mean transit time with the published transit times from two other epochs. The previous results come from the initial discovery data presented by \citet{gillon07b}, and the analyses of the \textit{Spitzer} light curve by G07, D07, and S08 that were discussed above. It should be noted that the time stamps of the \textit{Spitzer} photometry presented by G07, which was also analyzed by S08, are too late by 33\,s (M. Gillon private communication). The corrected central transit times for the G07 and S08 results are 2454280.78148 and 2454280.78174\,HJD respectively. We utilized these corrected times for all the invesigations described below.

The transit times determined by D07 and G07 are quite consistent, but disagree with the S08 determined value by more than expected from the given uncertainties ($\sim 2\sigma$). To compare the \textit{Spitzer} transit times with the other data we collapse the G07, D07, and S08 times into a single value by first averaging the G07 and S08 results, and then averaging this with the D07 result. We adopted the largest given uncertainty from the three analyses (0.00016\,d from D07) as the error in the collapsed transit times.

\begin{figure}
\resizebox{\hsize}{!}{\includegraphics{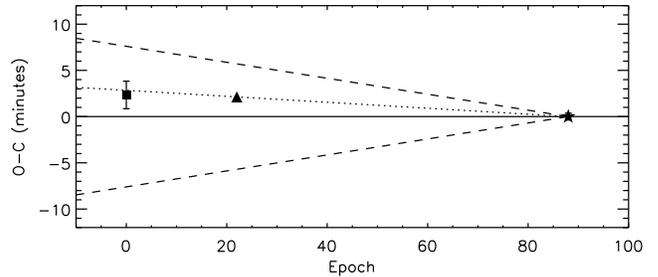}}
\caption{Observed transit time deviations from our ephemeris. The data are from \citet[][square]{gillon07b}, an average of the G07, D07, and S08 values (triangle), and this paper (star). For the later two values the error bars are smaller than the size of the points. The dashed lines give the $\pm 1\sigma$ range allowed by the uncertainty in the orbital period propagated backwards in time. The dotted line shows the trend expected due to the revised value of the period based on fitting the transit times.}
\label{f3}
\end{figure}

A plot of the observed transit time residuals from our ephemeris (transit time and orbital period) for the three available epochs is shown in Fig.~\ref{f3}. Our determined orbital period is consistent with, but 30\% more precise than, the value found by \citet{maness07} from analyzing the radial velocities of GJ\,436 only. We find a difference of 128\,s between the representative value from the \textit{Spitzer} observation and our own. The deviations show a nearly even trend with orbital epoch when also considering the lower precision time given by \citet{gillon07b}. This could be a result of an error in the period because the deviations are well within the range expected by the uncertainty in this parameter. Alternatively, the transit timing deviations could be due to perturbations of the host star arising from an additional object in the system.

\citet{maness07} found strong evidence for a long-term slope in the GJ\,436 radial velocities (1.4\,m\,s$^{-1}$\,yr$^{-1}$), which suggests it has an additional outer companion. The nature of this companion is currently unconstrained, but the observed long-term acceleration of the star means that it has a second degree of motion on top of its regular orbital motion due to the known transiting planet. Therefore, it is reasonable, and even expected, that the observed transit times of GJ\,436b are varying due to the changing position of the host star. What would be seen in such a case is a curvature in the transit timing offsets rather than the simple linear trend that would be due to an error in the orbital period.

Another possible way to detect perturbations with the transit timings is to look for deviations between the observed transit times and those predicted by orbital parameters found from modeling the radial velocities only. The key in this case is that the radial velocity orbit is relative to the system barycenter, which is invariant, and the transit times are relative to the position of the star, which should be changing. It is for this reason that we chose not to model our transit light curve simultaneously with the radial velocities. 

\citet{ribas08} have recently suggested that GJ\,436 hosts an additional ``Super-Earth'' planet in a near 2:1 resonance with the transiting planet. Such a planet would likely have a strong influence on the observed eclipse (both transit and secondary) times for planet b due to dynamical interactions. In this context, our finding a long-term trend in the transit timings could be a result of orbital precession caused by such interactions. Ultimately, a detailed dynamical study of the model proposed by \citet{ribas08} in comparison with the available eclipse times and radial velocities should be undertaken to evaluate their claim. This is beyond the scope of the current paper, but we plan to present such a study in the future.

If we assume that the transit timing deviations are due to an error in the orbital period then we may calculate a new, more precise value by leveraging the multiple orbits that have elapsed. Using our transit time and the average of the values reported by G07, D07, and S08 we find $P = 2.643904 \pm 0.000005$\,d, which is an order of magnitude more precise than the value we find by analyzing the \textit{HST} transit photometry alone. Adopting this period also brings the earlier transit time reported by \citet{gillon07b} in line with the other two, but this is not very significant due to the relatively low-precision of the value. After submitting this paper we became aware of recent transit times measured by \citet{alonso08} and amateur astronomers\footnote{See http://brucegary.net/AXA/GJ436/gj436.htm}. These data support our finding that the transit times for GJ\,436b are drifting from the \textit{Spitzer} ephemeris. The transit times we have determined will provide an additional benchmark for future studies aimed at detecting or setting limits on additional objects in the GJ\,436 system.

\begin{acknowledgements}
This work is based on observations made with the NASA/ESA \textit{Hubble Space Telescope}, obtained at the Space Telescope Science Institute, which is operated by the Association of Universities for Research in Astronomy, Inc., under NASA contract NAS 5-26555. These observations are associated with program GO/DD-11309. Support for this program was provided to J.B., G.B., B.M. and D.C. by NASA through a grant from the Space Telescope Science Institute. J.B., D.H., and S.D. acknowledge support from the DFG through grant number GRK 1351. A.S. and A.R. received funding from the DFG through grant number RE 1664/4-1. We thank the referee, Michael Gillon, for insightful comments that helped us improve this paper.

\end{acknowledgements}

\bibliographystyle{aa}
\bibliography{ms}

\end{document}